\documentclass[]{interact}

\usepackage{epstopdf}

\usepackage{mathtools}
\usepackage{amsmath,amssymb,amsthm, tabu}
\usepackage{optidef}
\usepackage{mathrsfs}
\usepackage{threeparttable}
\usepackage{float}
\usepackage{subcaption}
\usepackage{graphicx}
\usepackage[square,numbers]{natbib}

\theoremstyle{plain}

\theoremstyle{definition}

\theoremstyle{remark}

\theoremstyle{plain}

\DeclareMathOperator{\sign}{sign}

\begin{document}

\title{Observer-based Differentiators for Noisy Signals}

\author{
	\name{Van Thanh Huynh\textsuperscript{a}\thanks{CONTACT V.~T. Huynh. Email: v.huynh@deakin.edu.au}, Hieu Trinh\textsuperscript{a}, Riley Bain\textsuperscript{a}}
	\affil{\textsuperscript{a}School of Engineering, Deakin University, Australia;}
}

\maketitle

\begin{abstract}
We present a collection of different types of observation systems that work as differentiators. These observer-based differentiators can produce estimates for derivatives of a given signal, even though the given signal is prone to noise. 
\end{abstract}


\section{Motivation}

Differentiation of output measurements is essential for the realisation of unknown input observers discussed in this book. Many observer designs and control strategies rely on accurate estimates of signal derivatives to infer system states and unknown inputs. However, conventional numerical differentiation methods, such as finite differences, are highly sensitive to measurement noise—a common feature in practical applications. Even small, high-frequency noise components can be significantly amplified during differentiation, resulting in large errors and unstable derivative estimates. This sensitivity to noise presents a major challenge for systems that depend on reliable and robust derivative information.

To overcome these limitations, robust differentiation techniques have been developed in the literature, with observer-based approaches such as sliding mode and high gain observer differentiators gaining prominence. These methods are specifically designed to provide real-time, noise-resilient estimates of signal derivatives, ensuring stability and accuracy even in the presence of substantial measurement noise. By summarising and implementing these robust techniques, this chapter aims to equip practitioners and researchers with effective tools for accurate differentiation, thereby enhancing the performance and reliability of control and observation systems in noisy environments.
\section{Problem statement}
Let us consider a signal $\chi(t)$ as a function defined on $[0,\infty)$, which comprises $\delta(t)$ as a bounded Lebesgue-measurable noise with unknown features, and an unknown base signal $\chi_0(t)$. That means
\begin{align}
	\chi(t)=\chi_0(t)+\delta(t).
\end{align}

Assume that $\delta(t)$ is unknown but bounded, i.e. $|\delta(t)|\leq \varepsilon$. Since $\chi(t)$ is prone to noise, taking the direct first-order derivative of $\chi$ to infer $\dot{\chi}_0$ might lead to unstable and inaccurate computation of $\dot{\chi}_0(t)$. Instead, our aim is to obtain real-time, robust estimates of the first-order derivative of $\chi_0$ through the use of observation systems.

Let us denote by $x_1=\chi_0$, $x_2=\dot{\chi}_0$, and $\dot{x}_2=\ddot{\chi}_0=f(x,t)$. The first-order derivative process can be mathematically modelled as
\begin{align}
	\dot{x}_1&= x_2,\label{eqn:org_a}\\
	\dot{x}_2&=f(x,t),\label{eqn:org_b}\\
	y=x_1 +\delta(t) &= \chi(t) = \chi_0(t)+\delta(t),\label{eqn:org_c}
\end{align}
where $y$ is the measurement. Apparently, the measurement $y$ of this derivative process is also $\chi(t)$, which consists of the base signal $\chi_0(t)$ and noise $\delta(t)$.
 
The problem now becomes establishment of observers for system \eqref{eqn:org_a}-\eqref{eqn:org_c} so that we can robustly estimate in real time $x_1 = \chi_0$ and $x_2=\dot{\chi}_0$.
\section{Sliding mode differentiators}
Assuming that the unknown base signal has its first-order derivative with a known positive Lipschitz constant $L$. Sliding mode observers can be used as a differentiator \cite{Levant1998}. To develop the sliding mode observer for the derivative purpose, we introduce an observation system
\begin{equation}
	\dot{\hat{z}}_0(t) = v, \label{eqn:slm_aux}
\end{equation}
where $\hat{z}_0$ is used to estimate the base signal $\chi_0(t)$. Additionally, $v$ is a control input for the observation system. Let us call $s_0(t)=\hat{z}_0(t)-y(t)=\hat{z}_0(t)-\chi(t)$ the error between the estimate of the base signal and the original signal $\chi(t)$. We will design a control law $v$ using error $s_0$ so that $\hat{z}_0$ and $\dot{\hat{z}}_0=v$ from the observer provide estimates for $x_1=\chi_0$ and $x_2=\dot{\chi}_0$. 

Levant \cite{Levant1998,Levant2003} developed a dynamic, sliding mode control law $v$ based on the super-twisting principle that assures finite time convergence of $\hat{z}_0$ and $\dot{\hat{z}}_0$ towards $x_1=\chi_0$ and $x_2=\dot{\chi}_0$. The dynamic control law is proposed as follows:
\begin{align}
	v&=-\lambda_1|s_0|^{\frac{1}{2}}\sign(s_0)+\hat{z}_1,\label{eqn:slm_ctrl_a}\\
	\dot{\hat{z}}_1&=-\lambda_2\sign(s_0), \label{eqn:slm_ctrl_b}
\end{align}
where $\hat{z}_1$, which drives the control law $v$ in \eqref{eqn:slm_ctrl_a}, is derived from the newly-introduced dynamic model \eqref{eqn:slm_ctrl_b}. Additionally, $\lambda_1>0$ and $\lambda_2>0$ are tuning parameters. 

Observation system \eqref{eqn:slm_aux} under controller \eqref{eqn:slm_ctrl_a}-\eqref{eqn:slm_ctrl_b} works as a differentiator. Because $s_0=\hat{z}_0(t)-\chi(t)$, we have the sliding mode observer-based differentiator written in its full form as

\begin{align}
	\dot{\hat{z}}_0=v&=-\lambda_1|\hat{z}_0(t)-\chi(t)|^{\frac{1}{2}}\sign(\hat{z}_0(t)-\chi(t))+\hat{z}_1,\label{eqn:differentiator_slm_a}\\
	\dot{\hat{z}}_1&=-\lambda_2\sign(\hat{z}_0(t)-\chi(t)).\label{eqn:differentiator_slm_b}
\end{align}

This observer-based differentiator receives $y(t)=\chi(t)$ as its input and produces estimates $\hat{z}_0$, $\dot{\hat{z}}_0=v$ and $\hat{z}_1$. In fact, $\hat{z}_0(t)$ converges towards the base signal $\chi_0(t)$, while $\dot{\hat{z}}_0(t)=v$ serves as an estimate of the first-order derivative of the base signal $\dot{\chi}_0(t)$. 

When noise $\delta(t)$ is absent, $\hat{z}_0$ and $\dot{\hat{z}}_0=v$ asymptotically converge to $\chi_0(t)$ and $\dot{\chi}_0(t)$, as long as $\lambda_1$ is designed sufficiently large and $\lambda_2>L$. 

In the presence of noise, $\lambda_1$ and $\lambda_2$ can be tuned, for constants $\mu_1>0$ and $\mu_2>1$, so that
\begin{align}
\lambda_1&=\mu_1 L^{\frac{1}{2}}, ,\label{eqn:tuning_slm_a}\\
\lambda_2 &=\mu_2 L.\label{eqn:tuning_slm_b}
\end{align}
 
Again, the measurement $y$ is affected by noise $\delta(t)$ and that $|\delta(t)|\leq \varepsilon$. When $\lambda_1$ and $\lambda_2$ are tuned abiding \eqref{eqn:tuning_slm_a}--\eqref{eqn:tuning_slm_b}, the accuracy of the differentiator is proportional to $\varepsilon^{\frac{1}{2}}$. Specifically, there exists $b(\mu_1,\mu_2)>0$ such that
\begin{align}
	|v(t)-\dot{\chi}_0(t)|\leq b L^{\frac{1}{2}} \varepsilon^{\frac{1}{2}}.\label{eqn:est_error_bound}
\end{align}

According to \cite{Shtessel2014}, choosing $\lambda_1=1.5 L^{\frac{1}{2}}$ and $\lambda_2=1.1 L$ is recommended since this choice offers both a reasonably fast convergence and high accuracy.
 
We see that the upper-bound of the error between the base signal and the derivative generated by the sliding mode differentiator is affected by the Lipschitz constant of the base signal and the upper-bound of the noise. The Lipschitz constant also determines the design of the observer parameters $\lambda_1$ and $\lambda_2$. Therefore, prior knowledge about the Lipschitz constant of the first-order derivative of the base signal is necessary for the design of the sliding-mode differentiator.
   
\section{High-gain observer differentiators}
Assume that $f(x,t)$, which is the second-order derivative of $\chi_0(t)$, is locally Lipschitz in its arguments over a domain that contains the origin and  $f(0,0)=0$. Assume also that there exists $M>0$ such that

\begin{align}
|f(x,t)|\leq M.	
\end{align}

The differentiator can be implemented as a high-gain observer \cite{Khalil2008}. The high-gain observer differentiator is proposed by Khalil in \cite{Vasiljevic2006,Khalil2008,Khalil2017} as follows:

\begin{align}
	\dot{\hat{x}}_1 &= \hat{x}_2+\frac{\alpha_1}{\epsilon}(y - \hat{x}_1),\label{eqn:differentiator_hg_a}\\
	\dot{\hat{x}}_2 &= \frac{\alpha_2}{\epsilon^2}(y - \hat{x}_1), \label{eqn:differentiator_hg_b}
\end{align} 

where $\alpha_1$, $\alpha_2$, and $\epsilon$ are positive tuning parameters. Also, $\epsilon\ll 1$. The observation system is called a high-gain observer because observer gains $\alpha_1/\epsilon$ and $\alpha_2/\epsilon^2$ become extremely large as small $\epsilon$ moves towards $0$.
 
The high-gain observer provides estimates of $\chi(t)$ and $\dot{\chi}_0(t)$ via $\hat{x}_1(t)$ and $\hat{x}_2(t)$.

Tuning parameters $\alpha_1$ and $\alpha_2$ are designed so that polynomial $s^2+\alpha_1 s+\alpha_2$ is Hurwitz, i.e. real parts of the roots of the polynomial are strictly negative.

Convergence of the high-gain observer differentiator can be regulated by $\epsilon$. Faster convergence can be obtained when smaller $\epsilon$ is selected. However, very small $\epsilon$ will lead to an impulse-like transient behavior, which is called the \emph{peaking phenomenon}. 

Let $\tilde{x}_1(t)=x_1(t)-\hat{x}_1(t)$ and $\tilde{x}_2(t)=x_2(t)-\hat{x}_2(t)$ the estimation errors. Denote by $\tilde{x}=[\tilde{x}_1\;\tilde{x}_2]^T$. Moreover, $\tilde{x}_1(0)=x_1(0)-\hat{x}_1(0)$ and $\tilde{x}_2(0)=x_2(0)-\hat{x}_2(0)$ the initial estimation error. Also, let $\zeta=[\frac{\tilde{x}_1}{\epsilon}\; \tilde{x}_2]^T$. From system model \eqref{eqn:org_a}--\eqref{eqn:org_b} and high-gain observer model \eqref{eqn:differentiator_hg_a}--\eqref{eqn:differentiator_hg_b}, we can establish the error dynamics 
\begin{align}
	\dot{\zeta} = \frac{1}{\epsilon}A_\zeta \zeta - B f(x,t),
\end{align}
where 
\begin{align*}
	B &= \begin{bmatrix}
		0 & 1
	\end{bmatrix}^T,\\
	A_\zeta &= \begin{bmatrix}
		-\alpha_1 & 1\\
		-\alpha_2 & 0
	\end{bmatrix}.	
\end{align*}

Denote by $\bar{B}=\begin{bmatrix}
	-\alpha_1 & -\alpha_2
\end{bmatrix}^T$, $\Upsilon = \int_0^\infty{|\{e^{A_\zeta \tau}B\}_1|d\tau}$ and $\Phi = \int_0^\infty{|\{e^{A_\zeta \tau}\bar{B}\}_1|d\tau}$, $P = \int_0^\infty{|\{e^{A_\zeta \tau}B\}_2|d\tau}$ and $Q = \int_0^\infty{|\{e^{A_\zeta \tau}\bar{B}\}_2|d\tau}$, where $\{\bullet\}_k$ represents the $k$-th component of a vector. 

In the absence of noise $\delta(t)$, the estimation error is bounded by
\begin{align}
\vert\tilde{x}_2(t)\vert\leq c_1 + \epsilon^2 \Upsilon M,
\end{align}
where $c_1$ is an arbitrarily small positive constant.

Even if $\epsilon$ and $c_1$ are selected to be small, $c_1 + \epsilon^2 \Upsilon M>0$ is always a non-zero positive quantity. Thus, there is no asymptotic convergence to the base signals $\chi_0(t)$ and $\dot{\chi}_0(t)$ for the high-gain observer differentiator even when no noise exists. 

In the presence of noise $\delta(t)$ which is unknown and upperbounded by $\varepsilon>0$, Vasiljevic in \cite{Vasiljevic2006} indicated that the estimation error is bounded by
\begin{align}
	\vert\tilde{x}_2(t)\vert\leq c_1 + \epsilon P M +\frac{Q\varepsilon}{\epsilon}.\label{eqn:HGbound_noise}
\end{align}

In order to develop a tuning equation for $\epsilon$, let us consider the bound in \eqref{eqn:HGbound_noise}
\begin{align*}
	\eta(\epsilon) &=  c_1 + \epsilon P M +\frac{Q\varepsilon}{\epsilon},\\
	\frac{d\eta(\epsilon)}{d\epsilon} &= PM- Q\varepsilon\frac{1}{\epsilon^2},\\
	\frac{d^2\eta(\epsilon)}{d\epsilon^2} &= Q\varepsilon\frac{1}{\epsilon^3}>0,\\ 
\end{align*} 

We see that $\eta(\epsilon)$ is a convex function in $\epsilon$, and $\eta(\epsilon)$ has a minimum at 
\begin{align}
\epsilon^\star=\sqrt{\frac{Q\varepsilon}{PM}}. \label{eqn:tuning_hg_epsilon}
\end{align} 

Thus, the bound in \eqref{eqn:HGbound_noise} indicates that there is a balance threshold of $\sqrt{\frac{Q\varepsilon}{PM}}$ for $\epsilon$. Reducing $\epsilon$ below such a threshold will not result in improvement in the estimation errors. In fact, the ultimate bound $\eta(\epsilon)$ of the estimation errors increases when $\epsilon$ decreases further below $\sqrt{\frac{Q\varepsilon}{PM}}$. The high gain observer differentiator works better when $\frac{\varepsilon}{M}$ is relatively small so that a chosen $\epsilon$ can provide a tradeoff between fast convergence and attenuation of the effect of the noises. 

Equation \eqref{eqn:tuning_hg_epsilon} can be used to do the tuning of $\epsilon$.

\section{Simulation}
\subsection{Sliding mode differentiators}
\begin{figure}[h!] 
	\centering 
	\includegraphics[width=1.0\textwidth]{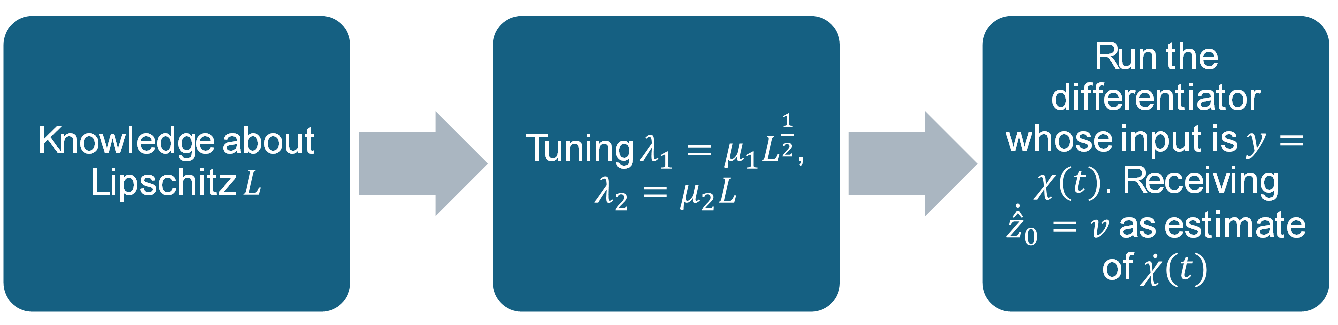}
	\caption{Steps of executing the sliding mode differentiators.}
	\label{fig:SLM_diagram} 
\end{figure}

Figure \ref{fig:SLM_diagram} shows steps that are recommended for tuning and executing the sliding mode differentiator. Since the tuning of differentiator parameters $\lambda_1$ and $\lambda_2$ depends on the Lipschitz constant $L$ of the first-order derivative of the unknown base signal, knowledge about $L$ is a prerequisite before using the sliding mode differentiator. 

The tuning of $\lambda_1$ and $\lambda_2$ then follows \eqref{eqn:tuning_slm_a}--\eqref{eqn:tuning_slm_b}. It is recommended in \cite{Shtessel2014} that $\lambda_1=1.5 L^{\frac{1}{2}}$ and $\lambda_2=1.1 L$ bring about a relatively fast convergence and accuracy.

We now simulate the sliding mode differentiator. Two simulation scenarios are conducted. One scenario is undertaken when there is no noise, while the other involves the presence of noise $\delta(t)$.

\begin{figure}[h!] 
	\centering 
	\includegraphics[width=1.0\textwidth]{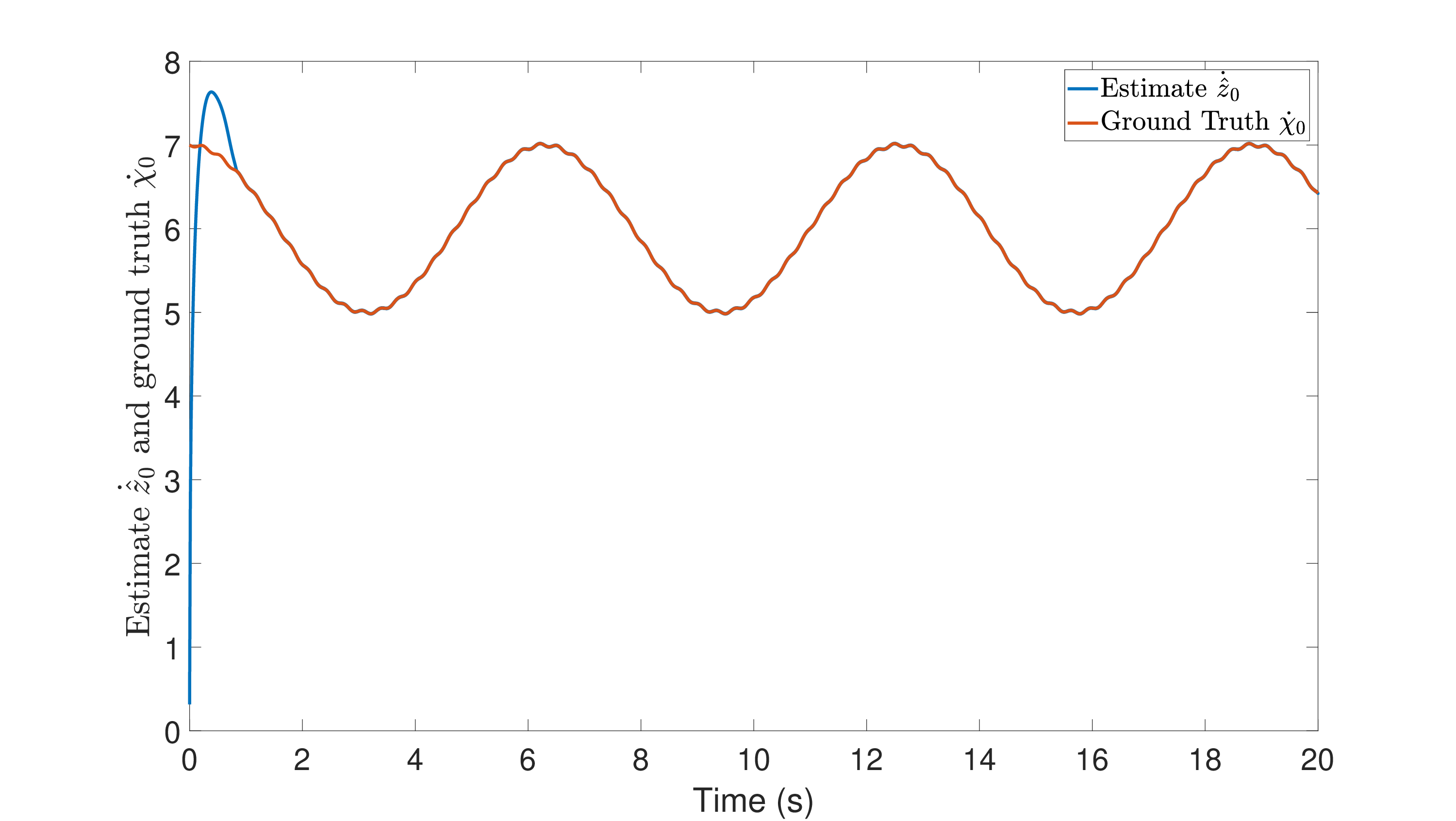}
	\caption{Asymptotic convergence of the estimate by the sliding mode differentiator when there is no noise.}
	\label{fig:SLM_noNoise} 
\end{figure}

In the first scenario of having no noise, the sliding mode observer (differentiator) is provided with a measurement $y(t)=\chi(t)=\chi_0(t)=6t+\sin(t)+0.001\cos(20t)$. On the one hand, taking the first order derivative of this noise-free measurement leads to a ground truth $\dot{\chi}_0(t)=6+\cos(t)-0.02\sin(20t)$. The ground truth is represented by the red line in Figure \ref{fig:SLM_noNoise}. On the other hand, the sliding mode differentiator gives an estimate of the first order derivative as $\dot{\hat{z}}_0=v$, which is represented by the blue line in Figure \ref{fig:SLM_noNoise}. It is apparent that the estimate produced by the sliding mode differentiator converges asymptotically to the ground truth when there is no noise. The sliding mode differentiator in this scenario is executed with $\lambda_1=10$  and $\lambda_2=8$ based on an assumed $L=6$. Such a tuning of $\lambda_1$ and $\lambda_2$ satisfies \eqref{eqn:tuning_slm_a}--\eqref{eqn:tuning_slm_b}, and simultaneously ensures $\lambda_1$ to be sufficiently large and $\lambda_2>L$ for the asymptotic convergence of the estimate.  

\begin{figure}[h!] 
	\centering 
	\includegraphics[width=1.0\textwidth]{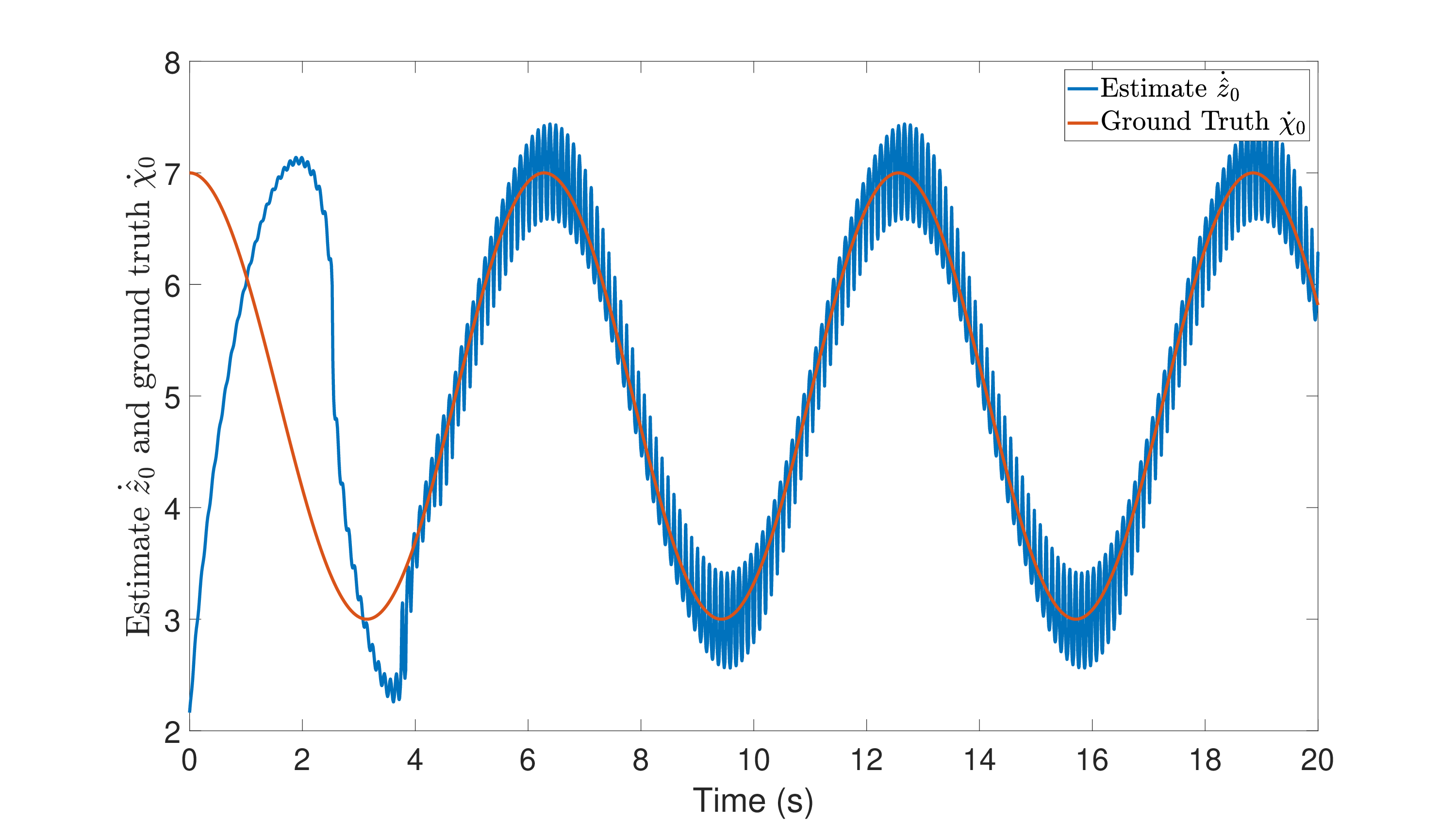}
	\caption{The estimate by the sliding mode differentiator when there is noise.}
	\label{fig:SLM_withNoise} 
\end{figure}

In the second scenario, the measurement is tarnished by noise $\delta(t)=0.04\cos(60t)$. Obviously, the noise is upper-bounded by $\varepsilon=0.04$. The base signal is $\chi_0(t)=5t+2\sin(t)+1$. Therefore, the input to the sliding mode differentiator is a measurement $y(t)=\chi(t)=\chi_0(t)+\delta(t)=5t+2\sin(t)+1+0.04\cos(60t)$. The ground truth is $\dot{\chi}_0=5+2\cos(t)$. Figure \ref{fig:SLM_withNoise} shows both the ground truth represented by the red line and the estimate produced by the sliding mode differentiator in the blue line. 

The observer in the second scenario is tuned with an assumption that $L=2$. Using \eqref{eqn:tuning_slm_a}--\eqref{eqn:tuning_slm_b}, where $\mu_1=1.5$ and $\mu_2=1.1$, results in $\lambda_1=2.12$ and $\lambda_2=2.20$. Moreover, the estimation error satisfies \eqref{eqn:est_error_bound}. In fact, for $b(\mu_1,\mu_2)=1.8$, the estimation error is proportional to $L^{\frac{1}{2}}\varepsilon^{\frac{1}{2}}$, and $|v(t)-\dot{\chi}_0(t)|\leq 1.8 \sqrt{2}\sqrt{0.04}=0.5091$.    
\subsection{High-gain observer differentiators}
\begin{figure}[h!] 
	\centering 
	\includegraphics[width=1.0\textwidth]{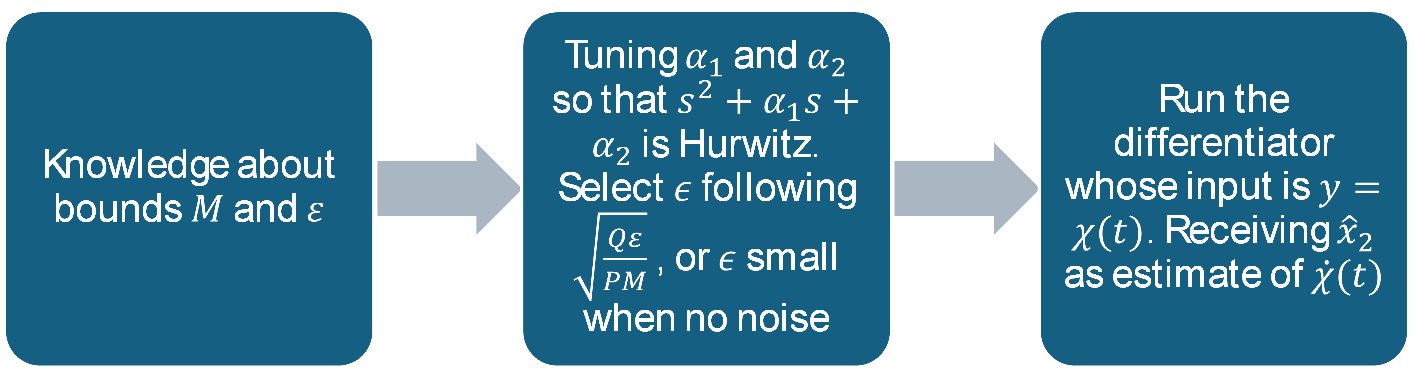}
	\caption{Steps of executing the high-gain differentiators influenced by noise.}
	\label{fig:HG_diagram} 
\end{figure}

Figure \ref{fig:HG_diagram} shows steps that are recommended for tuning and running the high-gain observer differentiator. Parameters $\alpha_1$ and $\alpha_2$ for the high-gain differentiator are chosen so that polynomial $s^2+\alpha_1 s+\alpha_2$ is Hurwitz. When there is no noise, parameter $\epsilon$ can be chosen small. When there is noise in the measurement output and knowledge of the noise upperbound is available, the tuning of $\epsilon$ depends on constant $M$ and upperbound $\varepsilon$. The tuning of $\epsilon$ then follows \eqref{eqn:tuning_hg_epsilon}. 

The high-gain observer differentiator is simulated in two scenarios, i.e. with the existence of noise and without the existence of noise. 

\begin{figure}[h!] 
	\centering 
	\includegraphics[width=1.0\textwidth]{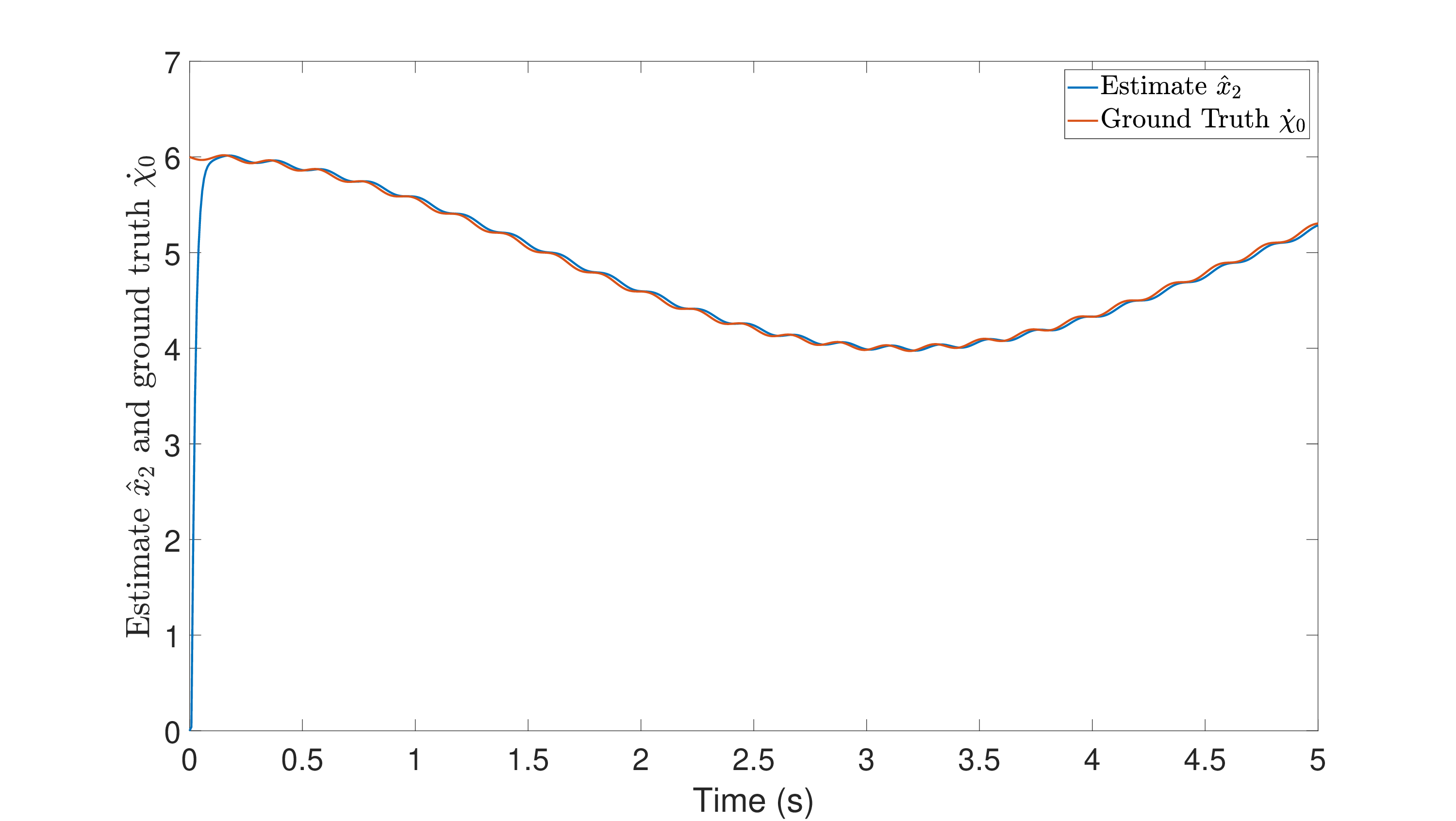}
	\caption{Convergence of the estimate by the high gain differentiator when there is no noise. The convergence is not asymptotic.}
	\label{fig:HG_noNoise} 
\end{figure}

In the noise-free scenario, the input to the high gain observer differentiator is a measurement $y(t)=\chi(t)=\chi_0(t)=5t+\sin(t)+0.001\cos(30t)$. Ground truth $\dot{\chi}_0(t)=5+\cos(t)-0.03\sin(30t)$ is simply obtained by mathematically taking the first order derivative of $\chi_0(t)$. Parameters of the high gain observer are $\alpha_1=6$ and $\alpha_2=8$, both of which make $s^2+\alpha_1 s+\alpha_2$ Hurwitz. For the sake of having small $\epsilon$, $\epsilon=0.04$ is selected. The red line in Figure \ref{fig:HG_noNoise} shows the ground truth. The blue line in Figure \ref{fig:HG_noNoise} represents $\hat{x}_2$ -- the estimate of $\dot{\chi}_0$ generated from the high-gain observer. It can be seen that estimate $\hat{x}_2$ converges closely to the ground truth even though the convergence is not asymptotic. 

In the scenario of having noise $\delta(t)=0.04\cos(60t)$, the simulated base signal is $\chi_0(t)=5t+\sin(t)+1$. Therefore, the input to the high-gain observer differentiator is $y(t)=\chi(t)=\chi_0(t)+\delta(t)=5t+\sin(t)+1+0.04\cos(60t)$, while the noise is upperbounded by $\varepsilon=0.04$. We can see that $M=1$. In addition, the ground truth is $\dot{\chi}_0(t)=5+2\cos(t)$.

The tuning of the parameters for the high gain observer in the presence of measurement noise follows the steps proposed in Figure \ref{fig:HG_diagram}. A selection of $\alpha_1=2$ and $\alpha_2=1$ makes $s^2+\alpha_1 s+\alpha_2$ Hurwitz. Moreover, calculating $\epsilon$ along \eqref{eqn:tuning_hg_epsilon} gives $\epsilon=0.1213$.

\begin{figure}[h!] 
	\centering 
	\includegraphics[width=1.0\textwidth]{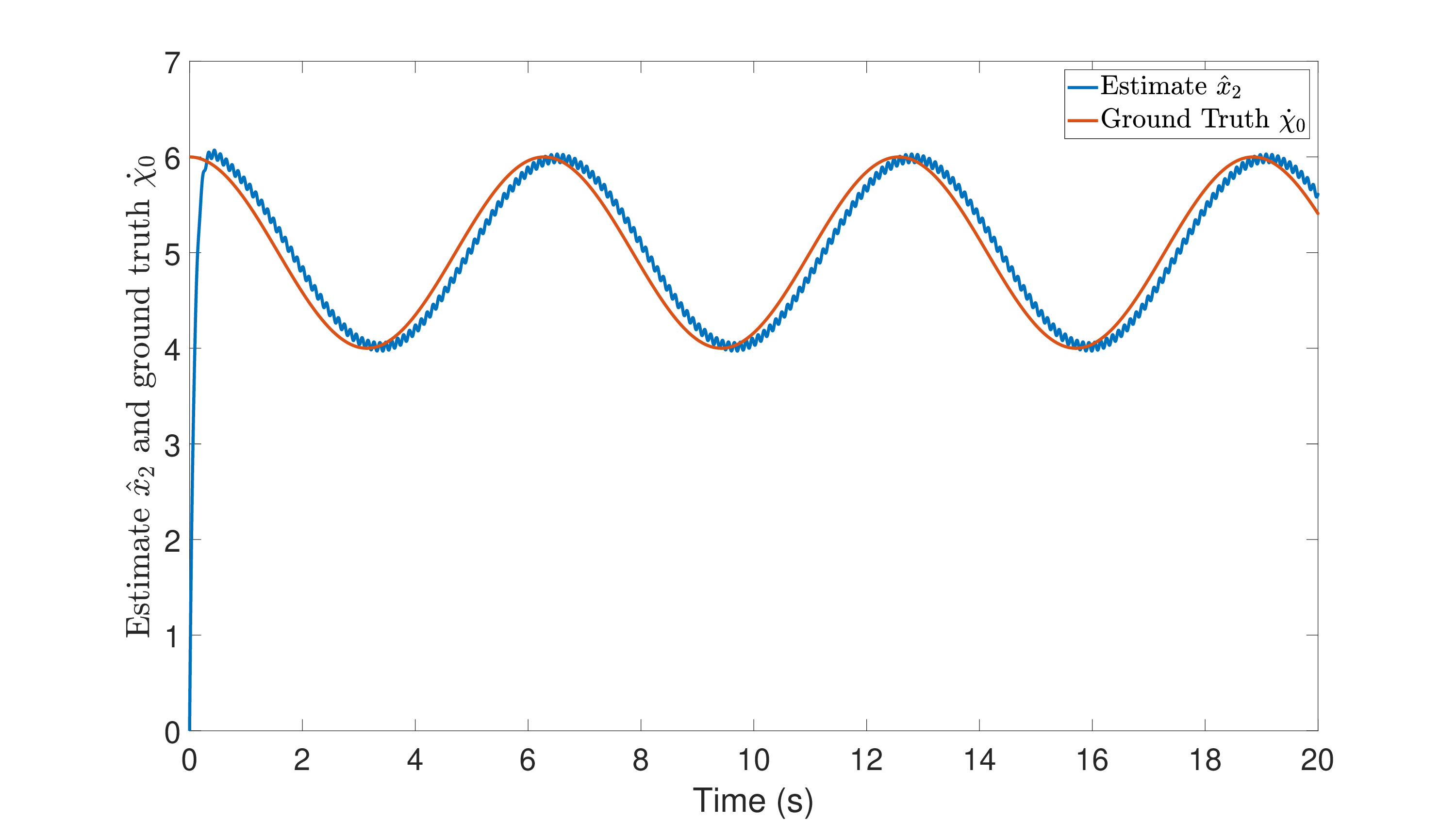}
	\caption{The estimate of the high gain observer differentiator when there is noise.}
	\label{fig:HG_withNoise} 
\end{figure}

The red line in Figure \ref{fig:HG_withNoise} shows the ground truth signal. The blue line in Figure \ref{fig:HG_withNoise} shows estimate $\hat{x}_2$ of $\dot{\chi}_0$. We see that the estimate converges towards the ground truth. Moreover, the estimation error is bounded. It can be verified that the bound of the estimation error in this simulation is less than $0.4852$, which stems from the computation of the bound using \eqref{eqn:HGbound_noise}.
\section{Concluding remarks}
In summary, this chapter has presented and analysed two robust observer-based differentiators—sliding mode and high-gain observer approaches—for estimating the first-order derivative of noisy signals. Through theoretical exposition and simulation, it is evident that both methods offer practical solutions for real-time differentiation in the presence of measurement noise, each with distinct tuning requirements and error bounds. The sliding mode differentiator demonstrates finite-time convergence and strong robustness, contingent on accurate knowledge of the Lipschitz constant, while the high gain observer provides a tunable trade-off between convergence speed and noise attenuation. These differentiators are essential tools for reliable derivative estimation in control and observation systems, and their careful implementation can significantly enhance the performance of systems subject to noisy measurements.

\bibliographystyle{plainnat}
\bibliography{./Bib/Citationshieu}     
\end{document}